\documentclass[11pt]{article} 

\usepackage[ansinew]{inputenc}
\usepackage{amsmath, amssymb, graphics, amsthm}
\usepackage{epsfig}
\usepackage{color}
\usepackage{fancyhdr} 

\usepackage[normalem]{ulem}

\newcommand{\D}{\underset{\leftarrow}}

\newcommand{\m}{\mbox{}}

\makeatletter
\@addtoreset{equation}{section}
\makeatother

\newcommand{\be}{\begin{equation}}
\newcommand{\ee}{\end{equation}}
\newcommand{\ba}{\begin{eqnarray}}
\newcommand{\ea}{\end{eqnarray}}

\def\pb#1{\rlap{\lower1.5ex\hbox{$\longleftarrow$}}{#1}}
\def\dpb#1{\rlap{\lower1.5ex\hbox{$\Longleftarrow$}}{#1}}
\def\spb#1{\rlap{\lower1.0ex\hbox{$\leftarrow$}}{#1}}
\def\sdpb#1{\rlap{\lower1.0ex\hbox{$\Leftarrow$}}{#1}}

\newcommand{\del}{\partial}

\title{{\sf The Wald entropy formula and loop quantum gravity}} 
\author{
{\sf N. Bodendorfer}$^{1,2}$\thanks{{\sf 
norbert@gravity.psu.edu, norbert.bodendorfer@fuw.edu.pl}},
{\sf Y. Neiman}$^{1,3}$\thanks{{\sf 
yashula@gmail.com}}
\\ {}\\
{\sf $^1$ Institute for
Gravitation and the Cosmos \& Physics
  Department,}\\
{\sf   Penn State, University Park, PA 16802, U.S.A.}\\ {}\\
{\sf  ${}^2$Faculty of Physics, University of Warsaw,}\\
{\sf   Pasteura 5, 02-093, Warsaw, Poland}\\ {}\\
{\sf  ${}^3$Perimeter Institute for Theoretical Physics,}\\
{\sf   31 Caroline St N, Waterloo, ON, N2L 2Y5, Canada}\\ 
}
\date{{\small\sf \today}}

\begin{document} 

\maketitle

{\sf

\begin{abstract}
We outline how the Wald entropy formula naturally arises in loop quantum gravity based on recently introduced dimension-independent connection variables. The key observation is that in a loop quantization of a generalized gravity theory, the analog of the area operator turns out to measure, morally speaking, the Wald entropy rather than the area. We discuss the explicit example of (higher-dimensional) Lanczos-Lovelock gravity and comment on recent work on finding the correct numerical prefactor of the entropy by comparing it to a semiclassical effective action. 
\end{abstract}

}

\section{Introduction}

A key feature that is expected from a theory of quantum gravity is an explanation for the thermodynamic behavior \cite{WaldTheThermodynamicsOf} of black holes observed in classical general relativity (GR). By today, several approaches to quantum gravity and semiclassical gravity have addressed this issue and offered different, at times seemingly unrelated, explanations. Moreover, the different approaches are not necessarily applicable to all classes of gravitational theories, such as Lanczos-Lovelock gravity or supergravity, or types of black hole solutions such as extremal or non-extremal black holes. It turns out, however, that the Wald entropy formula \cite{WaldBlackHoleEntropy}, applicable to general diffeomorphism invariant theories, agrees with other approaches where they are applicable. A derivation for the Wald entropy formula in the context of Euclidean quantum field theory has been given in \cite{BrownBlackHoleEntropy} for general diffeomorphism invariant theories. It is however desirable to understand the emergence of this general formula for the black hole entropy also from a more fundamental theory of quantum gravity.

Loop quantum gravity (LQG) \cite{RovelliQuantumGravity, ThiemannModernCanonicalQuantum} has emerged as a candidate theory for quantum gravity and addressed the question of black hole entropy with considerable success. See \cite{Diaz-PoloIsolatedHorizonsAnd} for a review. 
In LQG, one has been mainly interested in black hole entropy calculations for four-dimensional GR with minimally coupled matter fields. However, it was shown that for a non-minimally coupled scalar field, the black hole entropy can be obtained with the right dependence on the scalar field's value at the horizon \cite{AshtekarNonMinimallyCoupled, AshtekarNonMinimalCouplings, BSTI}, in accordance with the Wald formula. While this agreement presented an important confirmation for the robustness of the LQG framework, its proper origin remained obscure. 

In this paper, we will show how the Wald entropy formula naturally arises in LQG based on the recently introduced dimension-independent connection variables. The main idea is that for a generalized theory of gravity, such as Lanczos-Lovelock gravity or a non-minimally coupled scalar field, the direct analog of the area operator, which is a key ingredient in the entropy calculation, does not measure the area, but the Wald entropy. The reason is that the fluxes conjugate to the connection are given by derivatives of the Lagrangian with respect to certain curvature components, in the same way as in the construction of Wald entropy. Essentially, the canonical conjugate to the connection measures Wald entropy.

We present higher-dimensional Lanczos-Lovelock gravity as an explicit example. We shortly comment on general diffeomorphism invariant theories, where no robust general statements can be made from the LQG perspective at the moment. Our current understanding thus remains restricted to theories which can be formulated on the phase space of higher dimensional GR coupled to standard matter fields. Also, we restrict to theories which do not have additional constraints on top of the Hamiltonian and spatial diffeomorphism constraints of GR, plus additional gauge constraints or the simplicity constraint discussed below. Such additional constraints may require special treatment on top of the existing techniques in the LQG entropy calculations. Since an explicit solution for them might not be available, their effect on the black hole entropy would remain an open question. The class of treatable theories thus includes Lanczos-Lovelock gravity \cite{LanczosARemarkableProperty, LovelockTheEinsteinTensor} with non-minimal couplings of scalar fields, plus additional minimally coupled matter fields (the independence of the entropy on standard minimally-coupled matter fields is already well-understood \cite{AshtekarIsolatedHorizonsThe}).

The paper is organized as follows.
In section \ref{sec:preliminaries}, we review some background material concerning the Wald entropy formula, as well as the black hole entropy calculations in the isolated horizon framework in LQG. 
Next, we show in section \ref{sec:Lovelock} that the entropy calculation for higher-dimensional Lanczos-Lovelock gravity can be reduced to that of higher-dimensional GR. We explain how this result can be understood in general in terms of the Wald entropy formula. 
The prospects for general diffeomorphism-invariant theories are discussed in section \ref{sec:DiffInvTheory}.
In section \ref{sec:choices}, we comment on the interpretation of these results. Section \ref{sec:Conclusion} contains the conclusions of the paper.
In the appendix, we provide some details on the covariant phase space description omitted in section \ref{sec:Lovelock} and discuss the first law of black hole mechanics in Lanczos-Lovelock gravity within the isolated horizon framework.

\section{Preliminaries} \label{sec:preliminaries}

We begin with a brief review of the relevant fields. In section \ref{sec:preliminaries:wald}, we review the broad status of black hole entropy in generalized gravity theories. In sections \ref{sec:preliminaries:LQG_4d}-\ref{sec:preliminaries:LQG_higher}, we review the central features of the existing black hole entropy calculations in LQG. 

\subsection{Wald entropy} \label{sec:preliminaries:wald}

The black hole entropy formula for general diff-invariant theories of gravity was first proposed in \cite{WaldBlackHoleEntropy} and expanded on in \cite{IyerSomePropertiesOf,JacobsonOnBlackHole}. Previously, an entropy for the Lanczos-Lovelock class of theories was derived in \cite{JacobsonEntropyOfLovelock}. The defining property which motivated the original proposal \cite{WaldBlackHoleEntropy} was the First Law of thermodynamics for stationary black holes.  

A key concept in the entropy's derivation was that of a Noether potential \cite{WaldOnIdenticallyClosed}. For every local gauge symmetry of a field theory, there exists a Noether potential $Q^{\mu\nu}$. This is a rank-2 antisymmetric density, analogous to the Noether current $J^\mu$ that is associated with a global symmetry. While $J^\mu$ is integrated over a codimension-1 hypersurface to obtain a charge, $Q^{\mu\nu}$ is integrated over a codimension-2 surface. This is a generalization of the Gau{\ss} law from electromagnetism. In fact, the Noether potential for the electromagnetic gauge symmetry is just $\sqrt{-g}F^{\mu\nu}$; its integral through a codimension-2 spatial surface is the usual electric flux.

In \cite{WaldBlackHoleEntropy}, Wald used the Noether potential associated with diffeomorphisms, specifically with translations along the Killing field that becomes null at the event horizon. His insight was to view the First Law of black hole thermodynamics as a statement about the action variations under such translations. The expression for the entropy that arises from the First Law is then the integral of the Noether potential over the black hole's bifurcation surface. Now, bifurcation surfaces exist only for eternal stationary black holes. In \cite{JacobsonOnBlackHole}, it was shown that one can substitute the bifurcation surface by an arbitrary slice of a Killing horizon. This extended the applicability of Wald entropy to black holes that are only \emph{currently} stationary, having been formed in a dynamical process in the past.  

More explicitly, let the theory's Lagrangian be given by:
\begin{align}
 \mathcal{L} &= \mathcal{L} \left(g_{\mu\nu}, 
     R_{\mu\nu\rho\sigma}, \nabla_{\xi_1} R_{\mu\nu\rho\sigma}, \ldots, \nabla_{(\xi_1}\ldots \nabla_{\xi_n)} R_{\mu\nu\rho\sigma}, 
     \psi, \nabla_{\xi_1} \psi, \ldots, \nabla_{(\xi_l}\ldots \nabla_{\xi_l)} \psi \right) \ , \label{eq:L_general}
\end{align}
where $\psi$ denotes arbitrary matter fields, $\nabla_\mu$ is the covariant derivative associated with the metric $g_{\mu\nu}$, and $R_{\mu\nu\rho\sigma}$ is its Riemann curvature. Let $D+1$ be the spacetime dimension. The Wald entropy is then given by \cite{IyerSomePropertiesOf}:
\begin{align}
 S = -2\pi\int_H \sqrt{h}\, \frac{\delta\mathcal{L}}{\delta R_{\mu\nu\rho\sigma}} \epsilon_{\mu\nu}\epsilon_{\rho\sigma} d^{D-1}x \ . \label{eq:wald}
\end{align}
Here, $H$ is a slice of the horizon, $\sqrt{h}$ is its area density, $\epsilon_{\mu\nu}$ is its normal bivector with $\epsilon_{\mu\nu}\epsilon^{\mu\nu} = -2$, and $\delta\mathcal{L}/\delta R_{\mu\nu\rho\sigma}$ is the variational derivative of the Lagrangian with respect to $R_{\mu\nu\rho\sigma}$:
\begin{align}
 \begin{split}
   \frac{\delta\mathcal{L}}{\delta R_{\mu\nu\rho\sigma}}
    :={}& \frac{\del\mathcal{L}}{\del R_{\mu\nu\rho\sigma}} 
     - \nabla_{\xi_1} \left(\frac{\del\mathcal{L}}{\del\nabla_{\xi_1} R_{\mu\nu\rho\sigma}} \right) + \ldots \\
    &+ (-1)^n \nabla_{\xi_1}\ldots \nabla_{\xi_n} \left(\frac{\del\mathcal{L}}{\del \nabla_{(\xi_1} \ldots \nabla_{\xi_n)} R_{\mu\nu\rho\sigma}} \right)
    \ .
 \end{split} \label{eq:U}
\end{align}
In particular, for GR, one has:
\begin{align}
 \mathcal{L} = \frac{1}{16\pi G}R;\quad 
 \frac{\delta\mathcal{L}}{\delta R_{\mu\nu\rho\sigma}} = \frac{1}{32 \pi G}(g^{\mu\rho}g^{\nu\sigma} - g^{\mu\sigma}g^{\nu\rho});\quad
 S = \frac{1}{4G}\int_H \sqrt{h}\, d^{D-1}x = \frac{A_H}{4G} \ . \label{eq:Bekenstein}
\end{align}

The entropy formula \eqref{eq:wald} was also recovered in an analysis of the Euclidean black hole action \cite{BrownBlackHoleEntropy}, along the lines of the Gibbons-Hawking derivation \cite{GibbonsActionIntegralsAnd} for GR. A key difficulty was to properly handle the boundary contributions to the action, without the guidance of a standard variational principle which for GR leads to the York-Gibbons-Hawking boundary term. The solution was to notice that the conserved charges fix the boundary contribution at infinity, while at the bifurcation surface a nonvanishing contribution can only come from the extrinsic curvature.

Also notable is the detailed agreement between microscopic calculations of black hole entropy within string theory (first performed in \cite{StromingerMicroscopicOriginOf}) and the Wald formula. See e.g. \cite{CardosoCorrectionsToMacroscopic}. In this context, one applies the Wald formula to the effective supergravity action, with higher-derivative terms (tightly constrained by supersymmetry) arising from stringy effects. In some cases, the relevant part of the effective action can be found exactly, and the comparison between the Wald and microscopic entropies carried out to all orders in an asymptotic expansion \cite{DabholkarExactCountingOf}.
  
A major open question is how to define Wald entropy for non-stationary horizons. While the Bekenstein-Hawking area law \eqref{eq:Bekenstein} for GR extends unambiguously to a general horizon slice, that is not true for the more complicated formula \eqref{eq:wald}. This issue has been partially addressed in \cite{IyerSomePropertiesOf}. See also \cite{ChapmanFluidGravityCorrespondence} for a discussion in the context of black hole hydrodynamics. For Lanczos-Lovelock gravity, the imaginary action calculation in \cite{NeimanTheImaginaryPart}, as well as its Euclidean counterpart \cite{BanadosBlackHoleEntropy}, favor a particular non-stationary definition that depends only on the intrinsic geometry of the horizon slice. This result is in agreement with the more general proposal in \cite{IyerSomePropertiesOf}.

Other issues concern the positivity and increase properties of Wald entropy. Unlike the area law, the positivity of \eqref{eq:wald} may be field-dependent. This calls into question the general interpretation of \eqref{eq:wald} as a statistical entropy. Partial evidence suggests that negative Wald entropy is associated with pathological theories or solutions. In \cite{JacobsonBlackHoleEntropy}, it was shown that for a simple $f(R)$ theory, the positivity of Wald entropy is related to cosmic censorship. Also, in \cite{BrusteinWaldsEntropyIs}, it was suggested that the Wald formula can be viewed as the ordinary Bekenstein-Hawking entropy, but with the Newton's constant read off from the propagator of area perturbations on the horizon. This implies that the entropy's positivity may be correlated with the stability of the black hole.

Current understanding of entropy increase, i.e. the Second Law of thermodynamics, is again partial. In GR, the area increase theorem implies that the Second Law holds for very general processes. In general diff-invariant theories, it was shown \cite{JacobsonIncreaseOfBlack} that for \emph{quasi-stationary} accretion of matter, the Second Law is an automatic consequence of energy positivity. Little is known regarding non-stationary processes. One hope is that the requirement of entropy increase can serve to fix the correct \emph{definition} of entropy for non-stationary horizons. This has only been achieved \cite{JacobsonIncreaseOfBlack} for $f(R)$ theories, which are equivalent to GR with a non-minimally coupled scalar. In black hole hydrodynamics (a mildly non-stationary regime), the Second Law is expressed as a non-negative viscosity. In \cite{BriganteViscosityBoundViolation}, black hole hydrodynamics in Lanczos-Lovelock gravity was studied for planar black holes in 4+1d AdS space. It was shown there that whenever the AdS background exists, the viscosity is indeed non-negative. Finally, as possible evidence \emph{against} the Second Law, see the argument in \cite{LikoTopologicalDeformationOf} regarding entropy decrease in Lanczos-Lovelock gravity during black hole mergers.

\subsection{LQG entropy calculations in 3+1d} \label{sec:preliminaries:LQG_4d} 

The LQG calculation of black hole entropy has been originally performed in $3+1$ dimensions in terms of the Ashtekar-Barbero variables. The essential idea is that the canonical transformation from the ADM phase space \cite{ArnowittTheDynamicsOf} to Ashtekar-Barbero variables \cite{AshtekarNewVariablesFor, BarberoRealAshtekarVariables} yields a boundary contribution in the form of a Chern-Simons symplectic potential on the isolated horizon. The original calculations \cite{AshtekarIsolatedHorizonsThe, AshtekarMechanicsOfIsolated, AshtekarIsolatedHorizonsHamiltonian, AshtekarQuantumGeometryOf} were performed using a gauge fixing of SU$(2)$ to U$(1)$ on the isolated horizon. It was later shown that the calculation could be performed without this gauge fixing in a manifestly SU$(2)$-invariant manner \cite{EngleBlackHoleEntropy, EngleBlackHoleEntropyFrom}. 
The recent introduction of a higher-dimensional and supersymmetric generalization of loop quantum gravity \cite{BTTI,BTTII,BTTIII,BTTIV,BTTV,BTTVI,BTTVII} made it necessary to reconsider these calculations and extend them to the new framework \cite{BTTXII, BNI}. These results will be summarized in section \ref{sec:preliminaries:LQG_higher}. In the present subsection, we review the state of affairs in the original LQG framework, where the quantum aspects are well-developed. 

The basic idea of the entropy calculation is as follows. First, one derives from classical GR a boundary condition on the isolated horizon, along with a Chern-Simons symplectic potential. Upon quantization, this gives a quantum Chern-Simons theory describing the horizon degrees of freedom. In particular, the total area of the horizon is related to the Chern-Simons degrees of freedom by the quantum boundary condition. The dimension of the Chern-Simons Hilbert space, constrained to a fixed value of the total area, yields an entropy of the form: 
\begin{align}
 S = \frac{\alpha A_H}{\gamma G} \ . \label{eq:S_LQG}
\end{align}
Here, $\alpha$ is some numerical constant, while $\gamma$ is the Barbero-Immirzi parameter, a free parameter of the theory. It is tempting to set $\gamma = 4\alpha$ in order to obtain the well-known Bekenstein-Hawking entropy $A_H/4G$. However, this approach is probably too naive \cite{JacobsonANoteOn}. An intriguing alternative has recently been suggested \cite{FroddenBlackHoleEntropy}, involving an analytical continuation $\gamma\rightarrow\pm i$ in the large-spin limit. We defer these issues to section \ref{sec:choices}. 
For now, we remark that the area proportionality of the entropy is already a non-trivial result. In particular, it depends on using the correct combinatorics for the punctures, which follows from a proper study of the action of the diffeomorphism group.

Any derivation of black hole entropy should be tested on gravity theories with dynamics other than GR. Simple examples include Lanczos-Lovelock gravity \cite{LovelockTheEinsteinTensor}, as well as GR with non-minimally coupled matter, such as a conformally coupled scalar. It has been shown \cite{AshtekarNonMinimallyCoupled, AshtekarNonMinimalCouplings} that for the conformally coupled scalar, LQG produces the correct Wald-entropy analog of eq. \eqref{eq:S_LQG}. Specifically, one performs an LQG quantization using the Ashtekar-Barbero connection and its canonical conjugate (which is no longer the usual area flux). The standard isolated-horizon calculation then leads to the entropy formula:
\begin{align}
 S = \frac{\alpha A_H}{\gamma G} a(\phi); \quad a(\phi) = 1 - \frac{1}{6}\phi^2 \ . \label{eq:S_phi}
\end{align}
This is the Wald entropy for the conformally coupled scalar, up to the same constant factor as in \eqref{eq:S_LQG}.

In this paper, we aim to place the results \eqref{eq:S_LQG}-\eqref{eq:S_phi} in context, as well as to generalize them to higher-derivative gravity theories. We should note at this point that the result \eqref{eq:S_phi}, as well as our generalization of it, holds only for a particular choice of quantization variables. Specifically, the fundamental connection variable is taken to have the same geometric meaning as in ordinary LQG, so that the non-minimal couplings etc. affect only its conjugate flux. In section \ref{sec:choices}, we will discuss the implications and interpretation of different choices of variables. We will also comment there on how one should compare the LQG results for the entropy to the semiclassical Bekenstein-Hawking-Wald formula.

\subsection{Entropy in higher-dimensional LQG} \label{sec:preliminaries:LQG_higher}

In this paper, we consider an arbitrary number $D+1 \geq 3$ of spacetime dimensions. Up to the recent works \cite{BTTXII, BI, BII}, all papers on entropy calculations within LQG have been based on Ashtekar-Barbero or self-dual Ashtekar variables. In this subsection, we will briefly introduce the higher-dimensional connection variables, and review the entropy results \cite{BI, BII} that arise from them. 
Summarizing, it can be shown that the entropy computation in higher dimensions can be almost reduced to the well-studied $3+1$-dimensional case, up to the precise form of the area spectrum. This results from an isomorphism between the horizon Hilbert spaces in different dimensions, which has its origin in the implementation of the simplicity constraints \cite{BTTV}.

\subsubsection{Choice of bulk variables and constraints}

Let us give a little more detail and outline the results of \cite{BTTI, BTTII}. General Relativity in $D+1$ dimensions can be rewritten in terms of an SO$(D+1)$ Yang-Mills phase space. The conjugate variables are an SO$(D+1)$ connection $A_{aIJ}$ and a densitized generalized hybrid vielbein $\pi^{aIJ}$, related to the spatial metric via $2 q q^{ab} = \pi^{aIJ} \pi^b{}_{IJ}$, where $q = \det{q_{cd}}$. Here, $a,b=1,\ldots,D$ are spatial tensor indices on the spatial slice $\sigma$ in the $D+1$ decomposition of spacetime. $I,J=0,\ldots,D$ are internal indices transforming under SO$(D+1)$. In addition to the usual constraints of GR - a Hamiltonian constraint, a spatial diffeomorphism constraint and in our case an SO$(D+1)$ Gau{\ss} constraint - one must also introduce a so-called simplicity constraint. This constraint, given by $\pi^a_{[IJ} \pi^b_{KL]} \approx 0$, ensures\footnote{In four dimensions, a topological sector appears and the situation is more complicated, see \cite{BTTI}.} that $\pi^a_{IJ} = 2 n_{[I} E^a_{J]}$. Here, $n^I$ is a unit normal defined by $E^a_I n^I = 0$, while $E^a_I$ coincides in the ``time gauge'' $n^I = (1,0,\ldots,0)$ with the densitized $D$-bein derived from a spatial metric $q_{ab}$, i.e. $E^a_i E^b_j \delta^{ij} = q q^{ab}$, $i,j=1,\ldots,D$. Furthermore, an additional rescaling by a free real parameter $\beta$ takes place, so that the momentum becomes $ {}^{(\beta)} \pi^a_{KL} := \pi^a_{KL}/\beta$. This $\beta$ is analogous to, but different from the Barbero-Immirzi parameter $\gamma$. In the quantum theory, the simplicity constraint can be implemented\footnote{See \cite{BTTIII, BTTV} for a discussion of possible anomalies and the implementation on a node.} on the links of a spin-network by restricting the representations of SO$(D+1)$ to be of class $1$, so that their highest weight vector $\vec{\lambda}$ is determined by a single non-negative integer $\lambda$ as $\vec{\lambda} = (\lambda, 0, \ldots, 0)$ \cite{FreidelBFDescriptionOf}.

As in the standard isolated-horizon calculations (section \ref{sec:preliminaries:LQG_4d}), the canonical transformation to the variables above leads to a boundary term in the symplectic potential. 
This general phenomenon is related to the gauge invariance present in the theory and known as ``edge states'' in condensed matter physics, see \cite{BII} and references therein for more discussion on this point. As we will recall in the following, the resulting theory on the boundary can be written as a higher-dimensional Chern-Simons theory in even spacetime dimensions, and as a theory of bi-normals in general dimensions.
As usual, we take the boundary $H = \del\sigma$ of our hypersurface to be a horizon slice. 

\subsubsection{Even spacetime dimensions: Chern-Simons theory}

Let us denote the area density and the Euler density on $H$ by $\sqrt{h}$ and $E^{(D-1)}$, respectively. Now, restrict to a part of phase space where the scalar ratio $E^{(D-1)}/\sqrt{h}$ is constant (this is referred to in \cite{BTTXII} as the non-distortion condition; we will see in section \ref{sec:Lovelock} that it can be lifted). This condition implies in particular that up to a numerical factor, $E^{(D-1)}/\sqrt{h}$ is the same as $\chi/A_H$, where $\chi$ is the Euler characteristic of $H$. The second variation of the boundary symplectic potential can then be rewritten in terms of a Chern-Simons symplectic structure as \cite{BTTXII}:
\begin{align}
 \delta_{[1} \int_\sigma \del_a \left( \frac{1}{\beta} E^{aI} \delta_{2]} n_I\right) d^Dx 
  = \text{const} \times \frac{A_H}{\beta \chi} \int_H \, \text{Tr}_\epsilon 
   \left[ \delta_{[1}  \Gamma^0 \wedge \delta_{2]} \Gamma^0 \wedge R^0 \wedge \ldots \wedge R^0 \right] \ . \label{eq:SymplecticStructure}
\end{align}
Here, $\text{Tr}_\epsilon [X_1 X_2 \ldots X_{(D+1)/2}] := X_1^{IJ}  X_2^{KL} \ldots  X_{(D+1)/2}^{MN} \epsilon_{IJKL\ldots MN}$, $\Gamma^0$ is a generalization of the Peldan hybrid connection \cite{PeldanActionsForGravity} defined on $H$, and $R^0$ is the curvature of $\Gamma^0$. Furthermore, one can derive the boundary condition: 
\begin{align}
 \epsilon^{IJ \ldots KL MN} \epsilon^{\alpha \beta \ldots \delta \sigma} R^0_{\alpha \beta  IJ}  \ldots  R^0_{\delta \sigma KL}  = \text{const} \times  \frac{\beta \chi}{A_H} \times \hat s_a {}^{(\beta)}\pi^{aMN} \ ,
 \label{eq:BoundaryCondition}
\end{align}
where $\alpha,\beta=1,\ldots,D-1$ are tensor indices on $H$, $\hat s_a \equiv \epsilon_{a\beta\gamma\ldots\sigma}\epsilon^{\beta\gamma\ldots\sigma}/(D-1)!$ is the (appropriately densitized) normal covector to $H$ in $\sigma$, and $ {}^{(\beta)}\pi^{aMN} := \pi^{aMN} / \beta$. It seems that with such a Chern-Simons boundary theory in higher dimensions, a stronger form of the boundary condition is necessary to avoid having local degrees of freedom on the horizon \cite{BTTXII}. We will not discuss these details here, as they are not important for the arguments presented in this paper. The quantization sketched in section \ref{sec:Quantization} will be based on the bi-normals introduced in the following, and thus valid for any dimension $D+1 \geq 3$.

\subsubsection{General dimensions: Bi-normals}

Instead of using the right hand side of \eqref{eq:SymplecticStructure}, we can also use the left hand side as a definition of a Poisson bracket on $H$:
\be
	\{\tilde{s}^I(x), n_J(y)  \} = \beta  \, \delta^{I}_{J} \delta^{(D-1)}(x-y) \ ,	 \label{eq:BoundaryPoisson}
\ee
where $\tilde s_I \equiv E^a_I\hat s_a$ has unit density weight on $H$. The boundary condition 
\be
	2 / \beta \, n^{[I} \tilde s^{J]} = \hat s_a \pi^{aIJ} \label{eq:BoundaryConditionBiNormals}
\ee
now tells us that the proper variables to consider are given by $L^{IJ} = 2 / \beta \, n^{[I} \tilde s^{J]}$, that is the restriction of the fluxes to $H$. We can compute their Poisson algebra as 
\be
	\{ L_{IJ}(x), L_{KL}(y) \}  = 4 \,   \delta^{(D-1)}(x-y) \delta_{L][I} L_{J][K}(x)\text{.}  \label{eq:SympStrucL} 
\ee
The bi-normals have to be regularized like the fluxes in the bulk by smearing them over $D-1$-surfaces on $H$. The resulting Poisson algebra of smeared bi-normals is just the Lie algebra so$(D+1)$ at every puncture and thus agrees with algebra of the fluxes.

\subsubsection{Quantization}

\label{sec:Quantization}

A quantization based on the bi-normals as boundary variables has been given in \cite{BI}. Since the Poisson algebra of the bi-normals is just the Lie algebra so$(D+1)$, the representation problem is already solved and the boundary Hilbert space, before imposing any constraints, is a product of so$(D+1)$ representation spaces. By the boundary condition \eqref{eq:BoundaryConditionBiNormals}, the representation is only non-trivial at punctures.

Since the lapse function has to vanish at $H$, the Hamiltonian constraint does not have to be taken into account for the boundary Hilbert space \cite{AshtekarIsolatedHorizonsThe}. However, the remaining constraints, the spatial diffeomorphism constraint, the Gau{\ss} constraint, and the simplicity constraint, have to be dealt with. The spatial diffeomorphism constraint is solved in the standard way \cite{AshtekarQuantumGeometryOf}.
As discussed in \cite{BI, BII}, the Gau{\ss} law is implemented by projecting the boundary Hilbert space to its SO$(D+1)$ invariant subspace. The simplicity constraint restricts the intertwining representations labeling the invariant subspace in a certain recoupling scheme to be of class 1. By the results of \cite{BTTV}, the state counting problem is now reduced to the $3+1$-dimensional case \cite{EngleBlackHoleEntropy} based on SU$(2)$ Chern-Simons theory, since the boundary Hilbert spaces for a given set of puncture labels $\lambda_i$ have the same dimension when mapped as $\lambda_i = 2 j_i$.

From this point on, one can employ any of the strategies proposed in the literature for computing the black hole entropy. It is not our aim to provide an overview of these methods, or to advocate a particular one. We merely point out that the higher-dimensional computation can be reduced to the $3+1$-dimensional case, through the mapping of the boundary Hilbert spaces described above. Similarly, the main result of the present paper, detailed in section \ref{sec:Lovelock}, is that the computation in Lanczos-Lovelock gravity reduces to the one in general relativity.

\section{Entropy in Lanczos-Lovelock gravity from LQG} \label{sec:Lovelock}

\subsection{Canonical structure}

Lanczos-Lovelock gravity \cite{LovelockTheEinsteinTensor} is the most general higher-derivative theory of pure gravity that has no more than two time derivatives, so that it can be formulated on the same phase space as (higher-dimensional) GR. Up to a boundary term given in \cite{MyersHigherDerivativeGravity}, the Lanczos-Lovelock action reads:
\begin{align}
  S = \int_{\mathcal{M}} d^{D+1}x \, \sqrt{-g} \mathcal{L} 
   = \int_{\mathcal{M}} d^{D+1}x \, \sqrt{-g} \sum_{m=0}^{\lfloor \frac{D+1}{2} \rfloor } c_m \mathcal{L}_m \ , \label{eq:LovelockAction}
\end{align}
where $c_m$ are coupling constants, e.g. $c_1 = 1/(16\pi G)$, and: 
\begin{align}
  \mathcal{L}_m = \frac{(2m)!}{2^m} 
   R^{[\mu_1 \nu_1} {}_{[\mu_1 \nu_1} R^{\mu_2 \nu_2} {}_{\mu_2 \nu_2} \ldots R^{\mu_m \nu_m]} {}_{\mu_m \nu_m]}   \ .
\end{align}
The canonical formulation of this theory can be developed in analogy to the well-known ADM treatment \cite{ArnowittTheDynamicsOf}, and is given in \cite{TeitelboimDimensionallyContinuedTopological}. There is a certain problem in the analysis, since the extrinsic curvature cannot be expressed uniquely in terms of the metric and its canonical conjugate. This can lead to a multivalued Hamiltonian constraint. However, this does not seem to be of direct concern to us, since the constraint algebra and the spatial diffeomorphism constraint are unaffected \cite{TeitelboimDimensionallyContinuedTopological}. The Hamiltonian constraint does not enter the calculation, since the lapse function vanishes at the horizon \cite{AshtekarIsolatedHorizonsThe}. Maximally symmetric black hole solutions of the type considered in \cite{BTTXII} have been discussed in \cite{MaedaLovelockBlackHoles}  for higher-dimensional Lanczos-Lovelock gravity. The thermodynamics of Lanczos-Lovelock gravity has first been studied in \cite{MyersBlackholeThermodynamicsIn}, where it is shown that the entropy acquires a non-trivial prefactor depending on the coupling constants and the Riemann curvature of the horizon slice. 

Our goal in this section is to derive the canonical variables of Lanczos-Lovelock gravity that are relevant for the entropy calculation, and to relate them to the corresponding canonical variables in pure GR. 
We will omit some details about the phase space description that are not necessary for understanding the main point; these will be provided in the appendix.
It is more instructive to use the first-order Palatini formulation of the theory, since we want to calculate the momentum conjugate to the connection. Interestingly, Lanczos-Lovelock gravity is the only higher-derivative generalization of GR for which the first order Palatini-type action agrees with the second order Einstein-Hilbert-type action \cite{ExirifardLovelockGravityAt}. In particular, it follows from the field equations that the curvature of the connection is given by the Riemann tensor. This extends also to the vielbein-based first order formulations considered in this paper.

In first-order form, the Lanczos-Lovelock Lagrangian reads:
\begin{align}
  \mathcal{L} = -\sum_{m=0}^{\lfloor \frac{D+1}{2} \rfloor } c_m \frac{(2m)!}{2^m} \,
   e^{[\mu_1}_{I_1} e^{\nu_1}_{J_1} \ldots e^{\mu_m}_{I_m} e^{\nu_m]}_{J_m} \,
   F_{\mu_1 \nu_1} {}^{I_1 J_1} F_{\mu_2 \nu_2} {}^{I_2 J_2} \ldots F_{\mu_m \nu_m} {}^{I_m J_m} \ ,   \label{eq:LovelockActionFirst}
\end{align}
where $F_{\mu \nu} {}^{IJ}$ is the curvature of the SO$(1,D)$ connection $A_\mu{}^{IJ}$. The minus sign is chosen to agree with the conventions in \cite{BTTII}. The gauge group here is not SO$(D+1)$, since we are starting from a covariant framework. The transition from SO$(1,D)$ to SO$(D+1)$ as an internal gauge group while maintaining the Lorentzian signature of spacetime is detailed in \cite{BTTI, BTTII}, and involves a canonical transformation.
The reason that this trick works is based on the fact that both sets of connection variables are related via phase space reductions to an ADM-type phase space, which coincides for Euclidean and Lorentzian signature. The signature of spacetime is encoded in the Hamiltonian constraint of the theory, in case of GR as a relative sign between two terms. To avoid confusion, we will keep SO$(1,D)$ as the internal gauge group for the rest of this paper. The transition to SO$(D+1)$ would only change some signs.

The canonical momentum conjugate to $A_a^{IJ}$ reads:
\begin{align}
 \begin{split}
   \pi^a_{KL} &= e\frac{\del\mathcal{L}}{\del\dot A_a{}^{KL}} = - 2\sqrt{q}\, n_\mu \frac{\del\mathcal{L}}{\del F_{\mu a}{}^{KL}} \\
     &= 4\sqrt{q}\,n_\mu\sum_{m=1}^{{\lfloor \frac{D+1}{2} \rfloor}} m \frac{(2m)!}{2^m} c_m\,
      e^{[\mu}_K e^{a}_L e^{b_2}_{I_2} e^{c_2}_{J_2} \ldots e^{b_m}_{I_m} e^{c_m]}_{J_m}\,
      F_{b_2 c_2} {}^{I_2 J_2} \ldots F_{b_m c_m} {}^{I_m J_m} \ .
 \end{split} \label{eq:pi}
\end{align}
For $m=1$, we obtain the usual momentum $\pi^a_{IJ} = 2 n_{[I} E^a_{J]}$ with $E^a_J:=\sqrt{q} e^a_J$, familiar from the canonical analysis of the higher-dimensional Palatini action \cite{BTTII}. For consistency with \cite{BTTII}, we set $8\pi G = 1$, i.e. $c_1 = 1/2$. We note that in even spacetime dimensions, the topological (Gau{\ss}-Bonnet in $3+1$) contribution $c_{(D+1)/2} \neq 0$ does not affect the equations of motion, however it does change the canonical momentum, which will be important in what follows.

While we will show in the next paragraph that $\hat s_a \pi^{aIJ}$ is ``simple'' on $H$, that is it splits in the form $n^{[I} n_K \hat s_a \pi^{a|J]K}$ with $n^I = n^\mu e_\mu^I$, this is not true for $\pi^{aIJ}$ on generic phase space points. However, in order to apply the result of \cite{BTTI} that the canonical pair $\{A_{aIJ}, \pi^{bKL}\}$ can be reduced to the ADM-type canonical pair $\{\tilde q_{ab}, \tilde P^{cd}\}$ with $- 2 \tilde q \tilde q^{ab} = \pi^{aIJ} \pi^{b} {}_{IJ}$, we need that $\pi^{aIJ}$ is simple. 
This intermediate step is needed before a phase space extension to SO$(D+1)$ connection variables as in \cite{BTTI} can be performed.
Examining \eqref{eq:pi}, we see that a sufficient condition for this to hold is that $n^I F_{ab IJ} = 0$. On shell, this reduces to 
\be
	D_{[a} {K_{b]c}}=0 \text{,} \label{eq:ExtrinsicCurvatureCondition}
\ee
with $D_a$ being the covariant derivative with respect to $q_{ab}$ (the physical spatial metric) and $K_{ab}$ the extrinsic curvature. We will impose this condition as an operator in the quantum theory, and give more details on it in section \ref{sec:ExtrinsicCurvatureCondition}. With this restriction, we can treat Lovelock gravity in the same way as higher-dimensional general relativity and readily apply the results of \cite{BTTXII}. The question of whether the condition \eqref{eq:ExtrinsicCurvatureCondition} can be relaxed is left for future research.

We are interested in the effect of the new canonical momentum \eqref{eq:pi} on the black hole calculations. Thus, in analogy to \cite{AshtekarNonMinimallyCoupled, AshtekarNonMinimalCouplings, BSTI}, we must rewrite the boundary condition and the symplectic structure in terms of the new momentum. Here, several simplifications arise.
We are only interested in the canonical momentum on the horizon, and only in its $\hat s_a \pi^a_{IJ}$ component (recall that $\hat s_a$ is an appropriately densitized normal to $H$ within the hypersurface $\sigma$). This component reads:
\begin{align}
 \begin{split}
   \hat s_a\pi^a_{KL} &= -\sqrt{h}\, \epsilon_{\mu\nu} \frac{\del\mathcal{L}}{\del F_{\mu\nu}{}^{KL}} \\
    &= 2\sqrt{h}\,\epsilon_{\mu\nu} \sum_{m=1}^{{\lfloor \frac{D+1}{2} \rfloor}} m \frac{(2m)!}{2^m} c_m\,
      e^{[\mu}_{K} e^\nu_L e^{\alpha_2}_{I_2} e^{\beta_2}_{J_2} \ldots e^{\alpha_m}_{I_m} e^{\beta_m]}_{J_m}\,
      F_{\alpha_2 \beta_2} {}^{I_2 J_2} \ldots F_{\alpha_m \beta_m} {}^{I_m J_m} \ .
 \end{split} \label{eq:s_pi}
\end{align}
We see that all the field strengths in \eqref{eq:s_pi} are pulled back to $H$. It can be shown \cite{BTTXII} that this pullback equals $F_{\alpha\beta IJ} = R^0_{\alpha\beta IJ} = {}^{(D-1)}R_{\alpha \beta}{}^{\gamma \delta} e_{\gamma I} e_{\delta J}$, where ${}^{(D-1)}R_{\alpha \beta \gamma \delta}$ is the Riemann tensor of the intrinsic metric on $H$. It then follows that the $KL$ indices in \eqref{eq:s_pi} must lie in the plane of the binormal $\epsilon_{KL} = 2n_{[K}s_{L]}$. We get: 
\begin{align}
 \begin{split}
  \left.\hat s_a \pi^a_{KL} \right\vert_H &= \frac{1}{2}\sqrt{h}\,\epsilon_{KL} 
   \frac{\del\mathcal{L}}{\del F_{\mu\nu}{}^{IJ}}\epsilon_{\mu\nu}\epsilon^{IJ} \\
   &= \sqrt{h}\, \epsilon_{KL} \times \sum_{m=1}^{\lfloor \frac{D+1}{2} \rfloor} 2\,m\, c_m \frac{(2m-2)!}{2^{m-1}} ~ 
    {}^{(D-1)}R_{[\alpha_2 \beta_2} {}^{[\alpha_2 \beta_2} \ldots {}^{(D-1)}R_{\alpha_m \beta_m]} {}^{\alpha_m \beta_m]} \\
   &= -\sqrt{h}\,\epsilon_{KL} \frac{\del\mathcal{L}}{\del R_{\mu \nu \rho \sigma}} \epsilon_{\mu \nu}\epsilon_{\rho \sigma} \\
   &=: \sqrt{\tilde h}\,\epsilon_{KL} \ .
 \end{split} \label{eq:s_pi_long}
\end{align}
where $\epsilon_{\mu \nu} = 2n_{[\mu}s_{\nu]}$ is the binormal from section \ref{sec:preliminaries:wald}. Note that the equality between the first and third lines on the RHS of \eqref{eq:s_pi_long} follows from the equivalence between the first-order and second-order Lanczos-Lovelock actions. The factor $1/2$ between the first and the third line comes from different conventions for the definition of the derivative used in the literature: for the first line, we are consistent with \cite{BTTII}, while for the third line, we are consistent with \cite{WaldBlackHoleEntropy}. Again, the minus sign between these lines comes from the sign choice \cite{BTTII} for the first-order Lagrangian.

We see that $\hat s_a \pi^a_{KL}$ basically measures not the area density $\sqrt{h}$, but the density $\sqrt{\tilde h}$ of Wald entropy in units of $1/4G = 2\pi$ (this entropy density also has a geometric interpretation of sorts: roughly, the $m$-th order term is the topological Euler density in $2(m-1)$ dimensions times the area density in $D+1-2m$ dimensions). For the conformally coupled scalar field \cite{AshtekarNonMinimallyCoupled, AshtekarNonMinimalCouplings, BSTI}, we would have a similar result, with:
\begin{align}
 \hat s_a \pi^a_{KL} = \sqrt{\tilde h}\,\epsilon_{KL} 
  = -\sqrt{h}\,\epsilon_{KL} \frac{\del\mathcal{L}}{\del R_{\mu \nu \rho \sigma}} \epsilon_{\mu \nu}\epsilon_{\rho \sigma}
  = \sqrt{h}\,\epsilon_{KL}\left(1 - \frac{\phi^2}{6}\right) \ . 
\end{align}

Coming back to Lanczos-Lovelock theory, it remains to rewrite the boundary symplectic structure and the boundary condition from section \ref{sec:preliminaries:LQG_higher} using the Lanczos-Lovelock conjugate variables. 
We then find that \eqref{eq:SymplecticStructure} becomes:
\begin{align}
  \delta_{[1} \int_\sigma \partial_a \left( \frac{1}{\beta} \tilde{E}^{aI} \delta_{2]} n_I\right) d^Dx
   = \text{const} \times \frac{\tilde A_H}{ \chi \beta} \int_{H} \, \text{Tr}_{\epsilon} 
    \left[ \delta_{[1} \tilde{\Gamma}^0 \wedge \delta_{2]} \tilde{\Gamma}^0 \wedge \tilde{R}^0 \wedge \ldots \wedge \tilde{R}^0 \right] \ ,
\end{align}
Here, $\tilde A_H$ is the ``area'' given by the integral of $\sqrt{\tilde{h}}$, i.e. the Wald entropy in units of $1/4G = 2\pi$. The connection $\tilde{\Gamma}^0$ with curvature $\tilde{R}^0$, which in \cite{BTTXII} were built from the horizon metric, can now be built from {\it any} metric\footnote{More precisely, given a metric $\tilde h_{\alpha \beta}$ satisfying $\det (\tilde h_{\alpha \beta} )= \tilde h$, we construct a $D+1$-bein $\tilde e_{\alpha}^I$ such that $\tilde h_{\alpha \beta} = \tilde e_{\alpha}^I \tilde e_{\beta I}$ and $\tilde e_{\alpha}^I n_I = \tilde e_{\alpha}^I s_I = 0$. Then, $\tilde{\Gamma}^0 = \Gamma^0(\tilde e)$.} whose ``area'' density is $\sqrt{\tilde h}$. As in \cite{BTTXII}, the only additional restriction\footnote{It seems that a metric satisfying these two requirements can always be found for suitable horizon topologies: take a spherically symmetric metric $h^{s}_{ab}$ on $H$. Pick a diffeomorphism $\Phi$ on H such that $\sqrt{\tilde h} = \Phi^*(\sqrt{\det h^{s}_{\alpha \beta}})$. Then, $\tilde{h}_{\alpha \beta} :=\Phi^*( h^{s}_{\alpha \beta})$ also satisfies the non-distortion condition, since this condition is a scalar. Thus, the non-distortion condition does {\it not} pose any restriction on the actual spacetime metric. Still, $\chi = 0$ would lead to
ill-defined expressions in the Chern-Simons treatment and we exclude this particular case.} on this metric is that the associated Euler density $\tilde E^{(D-1)}$ satisfies $\tilde E^{(D-1)}/\sqrt{\tilde h} = \text{const}$ on $H$. As for the boundary condition \eqref{eq:BoundaryCondition}, it becomes:
\begin{align}
  \epsilon^{IJ \ldots KL MN} \epsilon^{\alpha \beta \ldots \delta \sigma} \tilde{R}^0_{\alpha \beta IJ}  \ldots  \tilde{R}^0_{\delta \sigma KL} 
   = \text{const} \times \frac{\chi\beta}{\tilde A_H} \times \hat s_a {}^{(\beta)}\pi^{aMN} \ .
\end{align}
The analog statement for the bi-normals is that $\tilde s^I$ and thus also $L^{IJ}$ are densitized with the Wald entropy density as opposed to the area density. 

\subsection{Entropy computation}

The quantization of \cite{BI}, as sketched in section \ref{sec:Quantization}, goes through as before, except the bulk states are now subject to the stronger condition \eqref{eq:ExtrinsicCurvatureCondition}, as we discuss in section \ref{sec:ExtrinsicCurvatureCondition}.
The only conceptual difference is that we are now counting states which correspond not to a given macroscopic area $A_H$, but to a given value of $\tilde A_H$. Indeed, the analog of the area operator built from the Lanczos-Lovelock fluxes has the familiar discrete spectrum \cite{BTTIII} $\text{const} \times \sqrt{\lambda (\lambda + D -2)}, ~ \lambda \in \mathbb{N}_0$, but measures $\tilde A_H$ rather than $A_H$. Using a straightforward generalization of the techniques developed in \cite{GhoshAnImprovedEstimate} to calculate the entropy (neglecting logarithmic corrections), one arrives at: 
\begin{align}
  S_{\text{Lovelock}} = \frac{\tilde\alpha \tilde A_H}{\beta G} \ , \label{eq:LovelockEntropy}
\end{align}
which is the correct Wald entropy up to a constant coefficient. $\tilde\alpha$ in \eqref{eq:LovelockEntropy} is a numerical constant analogous to $\alpha$ in \eqref{eq:S_LQG}, which depends on the number of dimensions.

We remark that the properties of isolated horizons used in \cite{BTTXII} remain valid in Lanczos-Lovelock gravity (and indeed in any theory), since they are of geometric origin and do not involve the field equations\footnote{\label{ftn:IHDefinition}This is not quite true given the definition in \cite{BTTXII} for pure GR. There, one imposes a version of the dominant energy condition on the energy-momentum tensor, which implies a similar condition on the Ricci tensor via the field equations. In a generalized theory, we must instead impose the condition directly on the Ricci tensor.}.

At this point, it becomes apparent why the Wald entropy formula and the LQG black hole entropy calculations agree: the generalized area operator is constructed roughly as $\tilde{A} \sim \sqrt{\text{flux}^2}$. The flux variables $ {}^{(\beta)} \pi^{a KL} $ conjugate to the connection are not measuring the (internal bivector-valued) area, but the derivative of the Lagrangian with respect to the curvature tensor component $R_{\mu\nu\rho\sigma}\epsilon^{\mu\nu}\epsilon^{\rho\sigma}\sim R_{nsns}$ as in the Wald entropy formula. Here, the first $n$ index comes from the time derivative of the connection when defining the conjugate momentum. The first $s$ index comes from the $s_a$-component of the momentum that's relevant for the entropy calculation. The second $n$ and $s$ result from the fact that for Lanczos-Lovelock gravity plus non-minimally coupled scalars, only the internal $s^{[I} n^{J]}$-component is non-vanishing. Whether this last statement is true in more general situations is presently unclear.

\subsection{Condition on extrinsic curvature} \label{sec:ExtrinsicCurvatureCondition}

Imposing \eqref{eq:ExtrinsicCurvatureCondition} at the classical level would lead us to use Dirac brackets, since this condition would gauge fix the Hamiltonian constraint. This would lead to a non-canonical symplectic structure, and the quantization methods of loop quantum gravity would cease to be applicable directly. Therefore, we choose to impose it at the quantum level in the form of a master constraint \cite{ThiemannQSD8}, which can also deal with second class constraints. The details of this treatment turn out not to be important for the following reason: using the standard regularization techniques, which lead to an anomaly free Hamiltonian constraint in the case of pure general relativity \cite{ThiemannQSD1}, the master constraint automatically vanishes on trivial (two-valent) vertices and extraordinary vertices (three-valent with two edges having parallel tangents). Therefore, the constraint acts non-trivially only in the bulk. Also, given any solution to the master constraint in the bulk with a certain set of spins puncturing the horizon, we can change the spins puncturing the horizon arbitrarily by adding edges and extraordinary vertices, and therefore generate another solution with arbitrary puncturing characteristics.

A few more remarks concerning the master constraint treatment should be given. Since we are dealing with second class constraints, we have to expect quantum corrections at the order of $\hbar$, as opposed to $\hbar^2$, comparable to a shift in energy, and corresponding to a classical reduction of degrees of freedom due to the constraints \cite{DittrichTestingTheMasterII}. In fact, imposing \eqref{sec:ExtrinsicCurvatureCondition} on top of the Hamiltonian constraint is a stronger condition on the bulk states than the usually assumed implementation of the Hamiltonian constraint. However, as remarked before, given a single solution, we can generate more solutions with arbitrary puncturing characteristics, which is sufficient to reduce the entropy computation to the usual case.
One might object to this procedure on the grounds that the solutions thus generated might dominate the entropy. A similar effect has been observed in \cite{DittrichTestingTheMasterV}, where volume zero vertices are ``overlooked'' in a master constraint treatment of the Gau{\ss} constraint. However, the same criticism can also be applied to the usual treatment, where one assumes a solution of the Hamiltonian constraint.

We remark that the Schwarzschild type black hole solutions \cite{MaedaLovelockBlackHoles} we are interested in satisfy condition \eqref{eq:ExtrinsicCurvatureCondition} in the standard time-independent slicing, since $K_{ab} = 0$ clearly implies \eqref{eq:ExtrinsicCurvatureCondition}. This special case corresponds to a time slice intersecting the bifurcation surface.

\section{The prospects for general diff-invariant theories} \label{sec:DiffInvTheory}

It was shown in \cite{BrownBlackHoleEntropy} that general diffeomorphism-invariant theories with Lagrangian of the form \eqref{eq:L_general} can be rewritten as a higher-dimensional gravity theory with no higher derivatives, coupled to additional (partially symmetric) tensor fields. Essentially, the additional degrees of freedom resulting from the higher time derivatives are traded for these tensor fields. In this process, new equations of motion which relate the tensor fields to derivatives of the Riemann tensor have to be imposed via Lagrange multipliers. In the canonical formalism, this translates into additional constraints.  

These results are an important step towards treating general diff-invariant theories within LQG, in the manner illustrated in section \ref{sec:Lovelock}. However, there remain several problems that prevent us from making any solid statements about LQG black hole entropy calculations for such theories. 
\begin{enumerate}

\item The canonical $D+1$ decomposition of symmetric tensor fields leads to additional terms proportional to the extrinsic curvature in the split action, and the calculation of the canonical conjugate to the extrinsic curvature becomes complicated. The expression of the split action given in \cite{BrownBlackHoleEntropy} hints at the conjugate of the extrinsic curvature being related to \eqref{eq:U}, thus potentially leading to an LQG derivation of the proper Wald entropy. However, this relies on treating the metric and extrinsic curvature as independent variables, which comes at the cost of additional second-class constraints. 

\item Symmetric tensor fields have not been treated so far by LQG methods (unlike $p$-forms - see \cite{BTTVII}). It seems that a construction similar to the connection variables for the metric might be necessary, which could yield additional boundary degrees of freedom. These degrees of freedom could contribute to the entropy.

\item After quantizing, one must take into account any leftover constraints. While the lapse function and thus the smeared Hamiltonian constraint vanish at the horizon, it is not clear what effects the additional first-class constraints might have on the entropy. 

\end{enumerate}

The situation for general diffeomorphism invariant theories is thus rather unclear at the moment. It seems that the best way to proceed is to study simple examples on a case by case basis to get a better feeling for them. We leave this for further research. 

\section{Interpretation and the choice of quantization variables} \label{sec:choices}

In the above, we've generalized the LQG entropy results \eqref{eq:S_LQG}-\eqref{eq:S_phi} to higher-derivative theories of gravity. We found that if one quantizes using the ordinary LQG connection (in the version appropriate to arbitrary dimensions, but otherwise retaining its geometric meaning) and its conjugate flux (which loses its simple geometric meaning), then the Wald entropy is recovered up to a constant factor. While this is an interesting result, it represents only a step towards a full understanding of black hole entropy within LQG. The caveats that must be raised appear already in the more familiar setting of LQG with Ashtekar-Barbero variables in 3+1d. Since more is known about that setting, we will use it for the purpose of the discussion. We expect that our comments below will also be relevant to the more general setup of sections \ref{sec:Lovelock}-\ref{sec:DiffInvTheory}.   

\subsection{Naive interpretation} \label{sec:choices:naive}

Let us begin with the standard LQG result \eqref{eq:S_LQG} for GR with minimally coupled matter. The naive response to this result is to set $\gamma = 4\alpha$, thus recovering by ``brute force'' the correct numerical coefficient for the Bekenstein-Hawking formula $S = A_H/4G$. Now, the Barbero-Immirzi parameter $\gamma$ (like the analogous parameter $\beta$ in the arbitrary-dimensional setup) defines a family of different quantization choices. Each is associated with a choice of fundamental connection variable, which is to be subjected to the LQG quantization procedure. Thus, the naive interpretation of \eqref{eq:S_LQG} would be that there is a single preferred choice of quantization variables. 

For GR with a conformally coupled scalar, this naive conclusion becomes sharper. One has now a larger selection of plausible connection variables to quantize. In particular, the constant parameter $\gamma$ can be replaced with a function of the scalar field $\phi$. Two choices appear especially natural. One choice, adopted in \cite{AshtekarNonMinimallyCoupled, AshtekarNonMinimalCouplings}, is to maintain the standard meaning of the fundamental connection, at the cost of its conjugate flux no longer measuring area. This is the choice described in section \ref{sec:preliminaries:LQG_4d} and the direct analog of the choice adopted by us in sections \ref{sec:Lovelock}-\ref{sec:DiffInvTheory}. It leads to the correct Wald entropy up to a constant, as shown in eq. \eqref{eq:S_phi}. Another choice\footnote{In \cite{BSTI}, on top of using the physical metric in the choice of canonical variables, a constant mean curvature gauge fixing of the Hamiltonian constraint was employed. This leads to fluxes rescaled by the scalar field, but by a different function from the one appearing in the Wald entropy. We will neglect this detail in this paper, as it is not important here, and the calculation in \cite{BSTI} can also be done without this gauge fixing.\label{ftn:NoCMC}} \cite{BSTI} is to maintain the geometric meaning of the fundamental flux as a measure of areas, at the cost of changing the meaning of the connection. This leads to the GR entropy formula \eqref{eq:S_LQG} instead of \eqref{eq:S_phi}, i.e. gives a wrong functional dependence of the entropy on the scalar field. Thus, again the naive conclusion is that there is a single preferred choice of quantization variables that produces the correct Wald entropy: 
\begin{enumerate}
 \item One must maintain the geometric meaning of the connection rather than the flux. This fixes the quantization variables up to a constant $\gamma$ and gives the correct Wald entropy up to a constant.
 \item Then, as in GR, one must fix further $\gamma = 4\alpha$. 
\end{enumerate}

As we will now review, this interpretation is in fact unfounded. On the other hand, a modified version of it appears to hold for large spins, as we will discuss in section \ref{sec:choices:large_spin}. 

\subsection{Semiclassical limits and the continuum} \label{sec:choices:classical}

The argument in section \ref{sec:choices:naive} is missing a crucial ingredient. One must always keep in mind that the Bekenstein-Hawking formula refers to a semiclassical regime of gravity. For instance, the Newton's constant $G$ appearing there comes from the prefactor of the semiclassical action. For Wald's generalization of the entropy formula, the same remark applies. Thus, any comparison of the LQG entropy to the Bekenstein-Hawking-Wald result must be in the context of some semiclassical limit. See \cite{JacobsonANoteOn} for a discussion. 

As discussed in \cite{BNI}, there are two semiclassical regimes that one may consider in LQG. One is the limit of continuum GR, which is supposed to emerge from LQG states with very many spins and intertwiners. If this limit exists, then it is of course the main focus of physical interest. If it doesn't, then the theory must be discarded as a description of nature. The continuum limit is also the supposed domain of the LQG entropy formula \eqref{eq:S_LQG}, since the latter receives contributions mainly from many small spins. However, we have no independent knowledge about the effective continuum action and its relation to the parameters of the fundamental theory. In particular, the relation between the fundamental Newton's constant (appearing in the entropy result \eqref{eq:S_LQG}) and the effective Newton's constant in the continuum (appearing in the Bekenstein-Hawking formula) is unknown. Thus, there is no direct conclusion that can be drawn from eqs. \eqref{eq:S_LQG}-\eqref{eq:S_phi} or from our generalization \eqref{eq:LovelockEntropy}. It may be that, as argued in section \ref{sec:choices:naive}, there is a unique quantization that correctly produces the continuum limit. Or it may be that all quantizations are equally good, and the results for the entropy are all correct when reexpressed in terms of the effective continuum action.

LQG has another semiclassical regime, which is not obviously related to the continuum one, but is much better understood technically. This is the limit of coherent states with very large spins -- a special subclass of states in the theory, characterized by large quantum numbers and large discrete elements of geometry. One then studies the contributions to the transition amplitudes where the intermediate states are again restricted to large spins. Since it isn't known that these contributions dominate, the large-spin regime may not be a proper limit of the theory at all. Nevertheless, its study -- the study of the spinfoam amplitudes at large quantum numbers -- has been fruitful.

With regard to black hole entropy, the situation that emerges in the large-spin regime is very different from the one in the continuum, with a conclusion quite similar (but not identical) to that of section \ref{sec:choices:naive}. It is the product of some surprising results \cite{FroddenBlackHoleEntropy,BNI} in standard, 3+1d LQG. These results are highly suggestive, but they are not yet rigorously understood from first principles\footnote{In particular, one would want to analytically continue the whole spin foam amplitude to derive the effective action, instead of only its asymptotic analysis.}. Also, at the moment, their applicability to higher dimensions and Lanczos-Lovelock gravity is only partial. However, we suspect that this new picture is important for a proper understanding of black hole entropy in LQG. We therefore describe it in the subsection below.

As an aside, the large-spin regime of LQG may also arise from a coarse-graining procedure; it would then be a legitimate rewriting of the continuum limit, rather than a mere truncation of the theory. However, this is far from guaranteed, and the theory's dynamics may be changed dramatically by RG flow. One favorable possibility is that LQG yields a continuum limit via a triangulation-invariant spinfoam model. In fact, if such a model were to arise from the RG flow of LQG, then we might as well take it as the fundamental theory, discarding the original one as scaffolding. If this model reproduces the GR action, then it may be similar to the familiar spinfoam amplitudes at large spins, since the latter reproduce the GR action as well (though the details of this are more subtle than previously thought -- see the next subsection).

\subsection{The large-spin limit and sending $\gamma$ to $\pm i$} \label{sec:choices:large_spin}

The recent calculation \cite{FroddenBlackHoleEntropy} by Frodden, Geiller, Noui and Perez suggests a new and intriguing perspective on the problem of black hole entropy in LQG. They show that for a fixed number of punctures with fixed large spins, after a certain analytical continuation, one obtains the Bekenstein-Hawking formula with the \emph{correct prefactor}. The analytical continuation is sending the Barbero-Immirzi parameter to a self-dual value $\gamma = \pm i$. At the same time, the puncture spins $j$ are sent to complex values, so that the puncture areas remain real.\footnote{We are referring here to the amended version of \cite{FroddenBlackHoleEntropy}, which makes the conceptual framework somewhat more solid. In the original version, it was the level of the Chern-Simons theory, rather than the spins, that became complex. The papers \cite{BSTI,BNI} were written referring to this original version. They are, however, fully compatible with the amended version. Note also that the authors of \cite{FroddenBlackHoleEntropy} interpret the analytically continued SU$(2)$ spins as labeling SU$(1,1)$ representations. This interpretation is not necessary for our argument.}
The analogous computation also works \cite{BSTI} for GR with a conformally coupled scalar, provided that one starts with the quantization variables from \cite{AshtekarNonMinimallyCoupled, AshtekarNonMinimalCouplings}, i.e. doesn't alter the geometric meaning of the connection. 

The use of a fixed number of large-spin punctures in \cite{FroddenBlackHoleEntropy} is very different from the usual approach, where all configurations are allowed, and small spins dominate. Accordingly, we are not suggesting that the calculation of \cite{FroddenBlackHoleEntropy} is directly relevant for the continuum limit. Instead, we view this result in the ``toy'' context of the large-spin semiclassical regime, which is distinct from the continuum as discussed in section \ref{sec:choices:classical}. 

The virtue of the large-spin regime is that we have a rather good understanding of its dynamics. In particular, we have at our disposal an effective action derived from spinfoam amplitudes. The one analyzed in the greatest detail is the 4-simplex vertex amplitude \cite{BarrettLorentzianSpinFoam}. Now, it was shown in \cite{BNI} that this amplitude reproduces the correct classical GR action only if one sets $\gamma = \pm i$ at the end of the calculation. This is contrary to previous claims that the correct action is reproduced for any real $\gamma$. The conclusion of \cite{BNI} rests on the recent observation \cite{NeimanOnShellActions, NeimanTheImaginaryPart} that the classical GR action has an imaginary part. A full agreement between the spinfoam amplitude and the classical action, \emph{including} the imaginary part, is obtained if and only if one sets $\gamma = \pm i$. 

In the same spirit as \cite{FroddenBlackHoleEntropy}, the continuation $\gamma = \pm i$ in \cite{BNI} is viewed as keeping the areas real, at the price of making the spins complex. Of course, it is not known how to define the quantum theory for non-real $\gamma$. At the moment, the closest one can get to a quantum theory based on self-dual variables is to send $\gamma\rightarrow\pm i$ \emph{after} the quantum calculation with real $\gamma$. This was the procedure used in \cite{BNI}. Real $\gamma$ can then be viewed as a regulator.

We note that the effective action of \cite{BNI, BarrettLorentzianSpinFoam} has the same Newton's constant as the fundamental theory. This is an expected result for coherent states with large quantum numbers, which need not hold for the continuum. The same Newton's constant also appears in the entropy result of \cite{FroddenBlackHoleEntropy}.

To sum up, the large-spin limit seems to describe a semiclassical regime only if one retains the geometric interpretation of the connection and sets $\gamma = \pm i$ after the quantum calculations have been performed. This procedure produces both a correct semiclassical action (i.e. a correct relation between its real and imaginary parts) and a correct black hole entropy (i.e. a correct relation between the entropy and the semiclassical action). 

So far in this subsection, we've been mostly reviewing the situation for $3+1$-dimensional general relativity. In higher dimensions and for Lanczos-Lovelock gravity, one can apply the entropy computation of \cite{FroddenBlackHoleEntropy} to the boundary Hilbert space from section \ref{sec:Quantization} \cite{BXV}. By the results of section \ref{sec:Lovelock}, the punctures will now carry quanta of the appropriate entropy-proportional quantity, and the method of \cite{FroddenBlackHoleEntropy} will produce the Wald entropy with the correct prefactor. However, as before, the entropy result is only meaningful when compared to an effective action, derived e.g. from a spinfoam amplitude. Writing a spinfoam model for generalized theories of gravity and calculating its large-spin behavior is a non-trivial task, and we will not comment on it further. 

\subsection{Quantized Wald entropy}

The expectation that black hole entropy has a discrete spectrum in quantum gravity has been put forward by several authors, starting with Bekenstein \cite{BekensteinTheQuantumMass}; see also \cite{BekensteinQuantumBlackHoles} for a review. A physical argument leading to this conclusion is to view black hole entropy as an adiabatic invariant, which is then expected to obtain a discrete spectrum in a quantum theory. Also, arguments for an equally spaced entropy spectrum can be found, see \cite{MedvedOnTheUniversal} and references therein. However, these arguments should be taken with care, since they are using semiclassical reasoning, and corrections in a deep quantum gravity regime are to be expected.  A generalization to Lanczos-Lovelock gravity can also be given \cite{KothawalaIsGravitationalEntropy}.

Comparing with the results of this paper, we recall that the area operator from standard loop quantum gravity had to be substituted by an operator that, when evaluated on a non-rotating isolated horizon, measures the Wald entropy. While its spectrum is not equidistant, it becomes nearly equidistant when evaluated on a single spin network edge for large quantum numbers. Such an edge is labeled by a simple representation of SO$(D+1)$, given by a non-negative integer $\lambda$ (see \cite{BTTIII, BTTV} for details). The eigenvalues are then given by $8 \pi G \beta \sqrt{\lambda(\lambda+D-1)}$, behaving as $8 \pi G \beta ( \lambda + (D-1)/2) + \mathcal O(1/\lambda)$ for $\lambda \rightarrow \infty$. The reason why the complete spectrum of this operator is not spaced equidistantly is this deviation from the equidistant spectrum for individual edges\footnote{See however \cite{BarberoFluxAreaOperator} for a different regularization of the area operator, resulting in an equidistant spectrum.}. 

In light of the present results, it is also interesting to revisit the interpretation of spin networks as ``twisted geometries'' \cite{FreidelTwistedGeometriesA}. This interpretation, made in the context of standard Einstein gravity, relies on the fact that the flux operator measures precisely the area. Then, the representation labels on the spin network edges can be interpreted as the magnitudes of the face areas of a polyhedron. However, in a general theory of gravity, we have seen that it is not the (spatial codimension 1) area that is quantized, but an expression proportional to the Wald entropy when evaluated on an isolated horizon. It is thus tempting to speculate that the microstates labeled by spin networks should have some interpretation which is closer to a collection of quantum black holes than to a discretized geometry. 

\section{Conclusion} \label{sec:Conclusion}

Working with the dimension-independent connection variables, we've related the LQG black hole entropy calculation to the Wald entropy formula. The key point is that the generalized area operators measure a rescaled version of the area at the horizon, which is essentially the Wald entropy. The reason for this is that the variable conjugate to the connection along the $\hat s_a$ direction (the spacelike horizon normal) is given by the derivative of the Lagrangian with respect to the curvature component $R_{nsns}$. The same quantity, integrated over the horizon slice $H$, enters the Wald entropy formula. Thus, the canonical conjugate to the connection essentially measures Wald entropy. Our analysis has covered non-minimally coupled scalars and Lanczos-Lovelock gravity.

The main open problem for the entropy calculation is the comparison with semiclassical actions. In \cite{BNI}, this has been done for four-dimensional pure gravity in a ``transplanckian'' large-spin regime. However, in the continuum, the problem remains open. The same is true even for large spins in higher dimensions, as well as for non-minimally coupled matter or Lanczos-Lovelock gravity, due to the lack of a corresponding spin foam model. 
As discussed in section \ref{sec:DiffInvTheory}, general diff-invariant theories are not yet under control. Reliable conjectures about their entropy as derived from loop quantum gravity, if such a calculation exists at all, cannot be made presently.

\section*{Acknowledgements}

NB and YN were supported by the NSF Grant PHY-1205388 and the Eberly research funds of The Pennsylvania State University. During final improvements of this work, NB was supported by a Feodor Lynen Research Fellowship of the Alexander von Humboldt-Foundation. Research at Perimeter Institute is supported by the Government of Canada through Industry Canada and by the Province of Ontario through the Ministry of Research \& Innovation. YN also acknowledges support of funding from NSERC Discovery grants. We thank the anonymous referee for constructive criticism and insightful comments, which helped to improve this paper.

\appendix

\section{First law and covariant phase space}
\label{sec:FirstLaw}

In this appendix, we fill in some details of the phase-space construction and the first law of black hole mechanics that were omitted in section \ref{sec:Lovelock}. We still restrict the phase space to satisfy \eqref{eq:ExtrinsicCurvatureCondition}, however point out that this condition might be relaxed in future work.

\subsection{Covariant phase space}

Following \cite{BTTXII}, we start with an action principle with suitable boundary terms, in our case the Lanczos-Lovelock action \eqref{eq:LovelockActionFirst}. We do not add any boundary term on the isolated horizon $\Delta$, since the isolated horizon boundary conditions are already ensuring a well-defined variational principle. In fact, under these conditions, the first variation of the action gets no boundary contribution from $\Delta$: 
\be
	\int_\Delta \tilde \Sigma^{IJ}  \wedge \delta A_{IJ} = 0 \text{,}
\ee
where $\tilde\Sigma^{IJ}$ is the $D-1$-form from which the flux operators are constructed:
\be
	\tilde \Sigma^{IJ} = \sum_{m=0}^{\lfloor \frac{D+1}{2} \rfloor} \frac{- m \, c_m}{2^{m-1} (D+1-2m)!} F^{I_2 J_2} \wedge \ldots \wedge F^{I_m J_m} \wedge e^{K_{2m+1}} \wedge \ldots \wedge e^{K_{D+1}} \epsilon_{IJI_2J_2 \ldots I_mJ_m K_{2m+1} \ldots K_{D+1}}.
\ee
The calculation is analogous to the one in \cite{BTTXII} and will not be detailed further here. Note that $\tilde \Sigma^{IJ}_{a_1 \ldots a_{D-1}} \epsilon^{a_1 \ldots a_D}  =  -\pi^{a_DIJ} / 2$. Next, one would like to calculate the second variation of the action, and show that the integral of the symplectic current over $\Delta$ reduces to a boundary term. This boundary term will constitute the horizon part of the symplectic structure. The required calculation is more involved than in the case of pure gravity \cite{BTTXII}, and we'll use a shortcut to circumvent it. Note that we are ultimately interested in switching the internal gauge group to $SO(D+1)$. For this, we have to go back to ADM-type variables and then apply the canonical transformation discussed in the previous section. In order to do this, we can pick a gauge where $n^I = \text{const.}$ and $s^I = \text{const.}$ (at least in individual charts) and thus $\delta n^I = \delta s^I = 0$. Then, the integral over $\Delta$ vanishes for the ADM-type variables, and we can perform the canonical transformation. In the covariant language, the symplectic structure now reads 
\be
	\Omega(\delta_1, \delta_2) =   \int_\Sigma \delta_{[1} \tilde \Sigma^{IJ} \wedge \delta_{2]} A_{IJ} + 2 \int_H \delta_{[1} \tilde s^I \delta_{2]} n_I \, d^{D-1}x \label{eq:SymplecticStuctureCovariant}
\ee
for the compact internal gauge group SO$(D+1)$.

\subsection{First law}

The derivation of the first law of black hole mechanics in the isolated horizon framework has been given in \cite{AshtekarIsolatedHorizonsHamiltonian}, and extended to higher dimensional black holes in anti-de Sitter spacetimes in \cite{AshtekarMechanicsOfHigher}. The main idea of the proof is to show that for an infinitesimal time translation to be a phase space symmetry, the first law of black hole mechanics must hold. The derivations in \cite{AshtekarIsolatedHorizonsHamiltonian, AshtekarMechanicsOfHigher} generalize in a straightforward manner to the case of Lanczos-Lovelock gravity. We will sketch here the important steps and perform the central calculation. We do not aim at being self-contained. The unfamiliar reader is referred to the original literature cited above for a detailed exposition, where also the inclusion of additional matter fields is discussed. 

Consider a time evolution vector field $t^\mu$. The variation $\delta_t := (\mathcal{L}_t e, \mathcal{L}_t A)$, where $\mathcal{L}_t$ is the Lie derivative along $t^\mu$, satisfies the linearized equations of motion for suitable boundary conditions, and can be interpreted as the generator of time evolution on the covariant phase space. However, $\delta_t$ constitutes a phase space symmetry only if $\mathcal{L}_t \Omega = 0$, where $\Omega$ is the symplectic structure. Since we are in the non-rotating case, the proper boundary condition for $t^\mu$ at the isolated horizon $\Delta$ is to become the null normal $l^\mu$. At spatial infinity, $t^\mu$ becomes an asymptotic time translation.

For the following calculation, we will use the fact that internal gauge transformations are already a symmetry of the symplectic structure. Therefore, we can assume without loss of generality that the variations $\delta$ in the following calculation do not contain gauge rotations, i.e. that $\delta n^I = \delta s^I = 0$, since $n^I$ and $s^I$ are normalized internal vectors. 
A tedious but straightforward calculation yields 
\ba
	\Omega(\delta, \delta_t) &=&  \frac{1}{2} \int_\Sigma \left( \delta \tilde \Sigma^{IJ} \wedge \mathcal{L}_t A_{IJ} - \mathcal{L}_t \tilde \Sigma^{IJ} \wedge \delta A_{IJ} \right)+ \int_H \left( \delta \tilde s^I \mathcal{L}_t  n_I - \mathcal{L}_t \tilde s^I \delta n_I \right) d^{D-1}x \nonumber \\
	&=& \frac{1}{2} \int_{\partial \Sigma} \left( \delta \tilde \Sigma^{IJ} \, t \cdot A_{IJ} - (t \cdot \tilde \Sigma^{IJ}) \wedge \delta A_{IJ}\right) + \int_{\Sigma} \underbrace{\left( \frac{1}{2} \delta F_{IJ} \wedge t \cdot \tilde \Sigma^{IJ} - (t \cdot F_{IJ}) \wedge \delta \tilde \Sigma^{IJ} \right)}_{ = \delta e^I \wedge (t \cdot (\text{Lanczos-Lovelock-EOM}))_I = 0} \nonumber \\
	&& + \int_H  \delta \tilde s^I \mathcal{L}_t n_I  \, d^{D-1}x \nonumber \\
	&=& -  \kappa^t \, \delta \int_H  \sqrt{\tilde h} \, d^{D-1}x+ \delta \int_{S_{\infty}} E^{t}_{\text{Lovelock}}\, d^{D-1}x := X_t(\delta)\text{,}
 \label{eq:FirstLaw}
\ea
where $\kappa^t$ is the surface gravity. In the last step, we have used that $t \cdot A_{IJ} = l^{\mu} \Gamma^0_{\mu IJ} - 2 \kappa n_{[I} s_{J]}$ on $H$ \cite{BTTXII} and the Lanczos-Lovelock equations of motion. Furthermore, the second term in the first integral of the second line vanishes when applying the isolated horizon boundary conditions \cite{BTTXII} to the connection and its curvature. $E^{t}_{\text{Lovelock}}$ is proportional to the generalization of the ADM energy density to Lanczos-Lovelock gravity. We refrain from writing out $E^{t}_{\text{Lovelock}}$ in detail, and just note that the corresponding term in \eqref{eq:FirstLaw} becomes a total variation due to the fall-off properties of the canonical variables at spatial infinity.
The first term in the last line of \eqref{eq:FirstLaw} evaluates to the expression $- \kappa^t \, \delta \tilde A_H$, familiar from the first law.
We are thus in the same situation as in \cite{AshtekarIsolatedHorizonsHamiltonian}. It remains to conclude that for the evolution to be Hamiltonian, $X_t(\delta)$ must be closed. This in turn implies that the surface gravity depends only on $\tilde A_H$, and that there exists a function $E^t_\Delta$ of $\tilde A_H$ such that $\delta E^t_\Delta = \kappa^t \, \delta \tilde A_H$. This $E^t_\Delta$ can then be interpreted as the horizon energy associated to the time translation $t^{\mu}$.
We refer to \cite{AshtekarMechanicsOfHigher, GravanisConservedChargesIn} for further details on this point. The results of \cite{AshtekarIsolatedHorizonsHamiltonian, AshtekarMechanicsOfHigher} thus generalize to Lanczos-Lovelock gravity in higher dimensions under the restriction \eqref{eq:ExtrinsicCurvatureCondition}\footnote{We remark that in order to lift the restriction \eqref{eq:ExtrinsicCurvatureCondition} for the first law, we would have to show in analogy to the pure GR calculation in \cite{BTTXII} that $\int_\Delta \tilde \Sigma^{IJ} \wedge \delta A_{IJ} = 0$ and $\int_{\Delta} \delta_{[1} \D{\tilde \Sigma}^{IJ} \wedge \delta_{2]} \D{A}\m_{IJ} = \int_{S_2} 2 (\delta_{[1} \tilde{s}^I)( \delta_{2]} n_I) -  \int_{S_1} 2 (\delta_{[1} \tilde{s}^I)( \delta_{2]} n_I)$, starting from \eqref{eq:LovelockActionFirst}. The tilde variables both refer to the Lovelock generalizations of those in \cite{BTTXII}. In the present paper, we used the above described shortcut of reducing to ADM-type variables to circumvent this computation by recycling the results of \cite{BTTXII}, leading to \eqref{eq:SymplecticStuctureCovariant}. This is in fact the only place where \eqref{eq:ExtrinsicCurvatureCondition} entered in this appendix.}. We leave the treatment of additional matter fields, as e.g. detailed in \cite{AshtekarIsolatedHorizonsHamiltonian}, to the interested reader.


\begin{thebibliography}{100}

\bibitem{WaldTheThermodynamicsOf}
R.~M. Wald, ``{The Thermodynamics of Black Holes},'' {\em Living Reviews in
  Relativity} {\bf 4} (2001) 6, {\tt arXiv:gr-qc/9912119}.

\bibitem{WaldBlackHoleEntropy}
R.~Wald, ``{Black hole entropy is the Noether charge},'' {\em Physical Review
  D} {\bf 48} (1993) R3427--R3431, {\tt arXiv:gr-qc/9307038}.

\bibitem{BrownBlackHoleEntropy}
J.~Brown, ``{Black hole entropy and the Hamiltonian formulation of
  diffeomorphism invariant theories},'' {\em Physical Review D} {\bf 52} (1995)
  7011--7026, {\tt arXiv:gr-qc/9506085}.

\bibitem{RovelliQuantumGravity}
C.~Rovelli, {\em {Quantum Gravity}}.
\newblock Cambridge University Press, Cambridge, 2004.

\bibitem{ThiemannModernCanonicalQuantum}
T.~Thiemann, {\em {Modern Canonical Quantum General Relativity}}.
\newblock Cambridge University Press, Cambridge, 2007.

\bibitem{Diaz-PoloIsolatedHorizonsAnd}
J.~Diaz-Polo and D.~Pranzetti, ``{Isolated Horizons and Black Hole Entropy in
  Loop Quantum Gravity},'' {\em SIGMA} {\bf 8} (2012) 048, {\tt arXiv:1112.0291
  [gr-qc]}.

\bibitem{AshtekarNonMinimallyCoupled}
A.~Ashtekar, A.~Corichi, and D.~Sudarsky, ``{Non-minimally coupled scalar
  fields and isolated horizons},'' {\em Classical and Quantum Gravity} {\bf 20}
  (2003) 3413--3425, {\tt arXiv:gr-qc/0305044}.

\bibitem{AshtekarNonMinimalCouplings}
A.~Ashtekar and A.~Corichi, ``{Non-minimal couplings, quantum geometry and
  black-hole entropy},'' {\em Classical and Quantum Gravity} {\bf 20} (2003)
  4473--4484, {\tt arXiv:gr-qc/0305082}.

\bibitem{BSTI}
N.~Bodendorfer, A.~Stottmeister, and A.~Thurn, ``{Loop quantum gravity without
  the Hamiltonian contraint},'' {\em Classical and Quantum Gravity} {\bf 30}
  (2013) 082001, {\tt arXiv:1203.6525 [gr-qc]}.

\bibitem{LanczosARemarkableProperty}
C.~Lanczos, ``{A Remarkable Property of the Riemann-Christoffel Tensor in Four
  Dimensions},'' {\em Annals of Mathematics} {\bf 39} (1938) 842--850.

\bibitem{LovelockTheEinsteinTensor}
D.~Lovelock, ``{The Einstein Tensor and Its Generalizations},'' {\em Journal of
  Mathematical Physics} {\bf 12} (1971), no.~3 498.

\bibitem{AshtekarIsolatedHorizonsThe}
A.~Ashtekar, A.~Corichi, and K.~Krasnov, ``{Isolated Horizons: the Classical
  Phase Space},'' {\em Advances in Theoretical and Mathematical Physics} {\bf
  3} (2000) 419--478, {\tt arXiv:gr-qc/9905089}.

\bibitem{IyerSomePropertiesOf}
V.~Iyer and R.~M. Wald, ``{Some properties of the Noether charge and a proposal
  for dynamical black hole entropy},'' {\em Physical Review D} {\bf 50} (1994)
  846--864, {\tt arXiv:gr-qc/9403028}.

\bibitem{JacobsonOnBlackHole}
T.~Jacobson, G.~Kang, and R.~Myers, ``{On black hole entropy},'' {\em Physical
  Review D} {\bf 49} (1994) 6587--6598, {\tt arXiv:gr-qc/9312023}.

\bibitem{JacobsonEntropyOfLovelock}
T.~Jacobson and R.~Myers, ``{Black hole entropy and higher curvature
  interactions},'' {\em Physical Review Letters} {\bf 70} (1993) 3684--3687,
  {\tt arXiv:hep-th/9305016}.

\bibitem{WaldOnIdenticallyClosed}
R.~M. Wald, ``{On identically closed forms locally constructed from a field},''
  {\em Journal of Mathematical Physics} {\bf 31} (1990) 2378.

\bibitem{GibbonsActionIntegralsAnd}
G.~W. Gibbons and S.~W. Hawking, ``{Action integrals and partition functions in
  quantum gravity},'' {\em Physical Review D} {\bf 15} (1977) 2752--2756.

\bibitem{StromingerMicroscopicOriginOf}
A.~Strominger and C.~Vafa, ``{Microscopic origin of the Bekenstein-Hawking
  entropy},'' {\em Physics Letters B} {\bf 379} (1996) 99--104, {\tt
  arXiv:hep-th/9601029}.

\bibitem{CardosoCorrectionsToMacroscopic}
G.~L. Cardoso, B.~de~Wit, and T.~Mohaupt, ``{Corrections to macroscopic
  supersymmetric black-hole entropy},'' {\em Physics Letters B} {\bf 451}
  (1999) 309--316, {\tt arXiv:hep-th/9812082}.

\bibitem{DabholkarExactCountingOf}
A.~Dabholkar, ``{Exact Counting of Supersymmetric Black Hole Microstates},''
  {\em Physical Review Letters} {\bf 94} (2005) 241301, {\tt
  arXiv:hep-th/0409148}.

\bibitem{ChapmanFluidGravityCorrespondence}
S.~Chapman, Y.~Neiman, and Y.~Oz, ``{Fluid/Gravity Correspondence, Local Wald
  Entropy Current and Gravitational Anomaly},'' {\em Journal of High Energy Physics} {\bf 1207} (2012) 128, {\tt arXiv:1202.2469 [hep-th]}.

\bibitem{NeimanTheImaginaryPart}
Y.~Neiman, ``{The imaginary part of the gravity action and black hole
  entropy},'' {\em Journal of High Energy Physics} {\bf 1304} (2013) 71, {\tt
  arXiv:1301.7041 [gr-qc]}.

\bibitem{BanadosBlackHoleEntropy}
M.~Ba\~{n}ados, C.~Teitelboim, and J.~Zanelli, ``{Black hole entropy and the
  dimensional continuation of the Gauss-Bonnet theorem},'' {\em Physical Review
  Letters} {\bf 72} (1994) 957--960, {\tt arXiv:gr-qc/9309026}.

\bibitem{JacobsonBlackHoleEntropy}
T.~Jacobson, G.~Kang, and R.~Myers, ``{BLACK HOLE ENTROPY IN HIGHER CURVATURE
  GRAVITY},'' {\tt arXiv:gr-qc/9502009}.

\bibitem{BrusteinWaldsEntropyIs}
R.~Brustein, D.~Gorbonos, and M.~Hadad, ``{Wald's entropy is equal to a quarter
  of the horizon area in units of the effective gravitational coupling},'' {\em
  Physical Review D} {\bf 79} (2009) 044025, {\tt arXiv:0712.3206 [hep-th]}.

\bibitem{JacobsonIncreaseOfBlack}
T.~Jacobson and G.~Kang, ``{Increase of black hole entropy in higher curvature
  gravity},'' {\em Physical Review D} {\bf 52} (1995) 3518--3528, {\tt
  arXiv:gr-qc/9503020}.

\bibitem{BriganteViscosityBoundViolation}
M.~Brigante, H.~Liu, R.~Myers, S.~Shenker, and S.~Yaida, ``{Viscosity bound
  violation in higher derivative gravity},'' {\em Physical Review D} {\bf 77}
  (2008) 126006, {\tt arXiv:0712.0805 [hep-th]}.

\bibitem{LikoTopologicalDeformationOf}
T.~Liko, ``{Topological deformation of isolated horizons},'' {\em Physical
  Review D} {\bf 77} (2008) 064004, {\tt arXiv:0705.1518 [gr-qc]}.

\bibitem{ArnowittTheDynamicsOf}
R.~Arnowitt, S.~Deser, and C.~W. Misner, ``{The dynamics of general
  relativity},'' in {\em Gravitation: An introduction to current research}
  (L.~Witten, ed.), (New York), pp.~227--265, Wiley, 1962.
\newblock {\tt arXiv:gr-qc/0405109}.

\bibitem{AshtekarNewVariablesFor}
A.~Ashtekar, ``{New Variables for Classical and Quantum Gravity},'' {\em
  Physical Review Letters} {\bf 57} (1986) 2244--2247.

\bibitem{BarberoRealAshtekarVariables}
J.~Barbero, ``{Real Ashtekar variables for Lorentzian signature space-times},''
  {\em Physical Review D} {\bf 51} (1995) 5507--5510, {\tt
  arXiv:gr-qc/9410014}.

\bibitem{AshtekarMechanicsOfIsolated}
A.~Ashtekar, C.~Beetle, and S.~Fairhurst, ``{Mechanics of isolated horizons},''
  {\em Classical and Quantum Gravity} {\bf 17} (2000) 253--298, {\tt
  arXiv:gr-qc/9907068}.

\bibitem{AshtekarIsolatedHorizonsHamiltonian}
A.~Ashtekar, S.~Fairhurst, and B.~Krishnan, ``{Isolated horizons: Hamiltonian
  evolution and the first law},'' {\em Physical Review D} {\bf 62} (2000)
  104025, {\tt arXiv:gr-qc/0005083}.

\bibitem{AshtekarQuantumGeometryOf}
A.~Ashtekar, J.~Baez, and K.~Krasnov, ``{Quantum Geometry of Isolated Horizons
  and Black Hole Entropy},'' {\em Advances in Theoretical and Mathematical
  Physics} {\bf 4} (2000) 1--94, {\tt arXiv:gr-qc/0005126}.

\bibitem{EngleBlackHoleEntropy}
J.~Engle, K.~Noui, and A.~Perez, ``{Black Hole Entropy and SU(2) Chern-Simons
  Theory},'' {\em Physical Review Letters} {\bf 105} (2010) 031302, {\tt
  arXiv:0905.3168 [gr-qc]}.

\bibitem{EngleBlackHoleEntropyFrom}
J.~Engle, K.~Noui, A.~Perez, and D.~Pranzetti, ``{Black hole entropy from an
  SU(2)-invariant formulation of Type I isolated horizons},'' {\em Physical
  Review D} {\bf 82} (2010) 044050, {\tt arXiv:1006.0634 [gr-qc]}.

\bibitem{BTTI}
N.~Bodendorfer, T.~Thiemann, and A.~Thurn, ``{New variables for classical and
  quantum gravity in all dimensions: I. Hamiltonian analysis},'' {\em Classical
  and Quantum Gravity} {\bf 30} (2013) 045001, {\tt arXiv:1105.3703 [gr-qc]}.

\bibitem{BTTII}
N.~Bodendorfer, T.~Thiemann, and A.~Thurn, ``{New variables for classical and
  quantum gravity in all dimensions: II. Lagrangian analysis},'' {\em Classical
  and Quantum Gravity} {\bf 30} (2013) 045002, {\tt arXiv:1105.3704 [gr-qc]}.

\bibitem{BTTIII}
N.~Bodendorfer, T.~Thiemann, and A.~Thurn, ``{New variables for classical and
  quantum gravity in all dimensions: III. Quantum theory},'' {\em Classical and
  Quantum Gravity} {\bf 30} (2013) 045003, {\tt arXiv:1105.3705 [gr-qc]}.

\bibitem{BTTIV}
N.~Bodendorfer, T.~Thiemann, and A.~Thurn, ``{New variables for classical and
  quantum gravity in all dimensions: IV. Matter coupling},'' {\em Classical and
  Quantum Gravity} {\bf 30} (2013) 045004, {\tt arXiv:1105.3706 [gr-qc]}.

\bibitem{BTTV}
N.~Bodendorfer, T.~Thiemann, and A.~Thurn, ``{On the implementation of the
  canonical quantum simplicity constraint},'' {\em Classical and Quantum
  Gravity} {\bf 30} (2013) 045005, {\tt arXiv:1105.3708 [gr-qc]}.

\bibitem{BTTVI}
N.~Bodendorfer, T.~Thiemann, and A.~Thurn, ``{Towards loop quantum supergravity
  (LQSG): I. Rarita-Schwinger sector},'' {\em Classical and Quantum Gravity}
  {\bf 30} (2013) 045006, {\tt arXiv:1105.3709 [gr-qc]}.

\bibitem{BTTVII}
N.~Bodendorfer, T.~Thiemann, and A.~Thurn, ``{Towards loop quantum supergravity
  (LQSG): II. p -form sector},'' {\em Classical and Quantum Gravity} {\bf 30}
  (2013) 045007, {\tt arXiv:1105.3710 [gr-qc]}.

\bibitem{BTTXII}
N.~Bodendorfer, T.~Thiemann, and A.~Thurn, ``{New variables for classical and
  quantum gravity in all dimensions: V. Isolated horizon boundary degrees of
  freedom},'' \\{\em Classical and Quantum Gravity} {\bf 31} (2014) 055002, {\tt
  arXiv:1304.2679 [gr-qc]}.

\bibitem{BNI}
N.~Bodendorfer and Y.~Neiman, ``{Imaginary action, spinfoam asymptotics and the
  'transplanckian' regime of loop quantum gravity},'' {\em Classical and
  Quantum Gravity} {\bf 30} (2013) 195018, {\tt arXiv:1303.4752 [gr-qc]}.

\bibitem{JacobsonANoteOn}
T.~Jacobson, ``{A note on renormalization and black hole entropy in loop
  quantum gravity},'' {\em Classical and Quantum Gravity} {\bf 24} (2007)
  4875--4879, {\tt arXiv:0707.4026 [gr-qc]}.

\bibitem{FroddenBlackHoleEntropy}
E.~Frodden, M.~Geiller, K.~Noui, and A.~Perez, ``{Black Hole Entropy from
  complex Ashtekar variables},'' 
  {\em Europhysics Letters} {\bf 107} (2014) 10005, 
  {\tt arXiv:1212.4060 [gr-qc]}.

\bibitem{BI}
N.~Bodendorfer, ``{Black hole entropy from loop quantum gravity in higher
  dimensions},'' {\em Physics Letters B} {\bf 726} (2013) 887--891, {\tt
  arXiv:1307.5029}.

\bibitem{BII}
N.~Bodendorfer, ``{A note on entanglement entropy and quantum geometry},'' \\{\em Classical and Quantum Gravity} {\bf 31} (2014) 214004, {\tt
  arXiv:1402.1038 [gr-qc]}.

\bibitem{FreidelBFDescriptionOf}
L.~Freidel, K.~Krasnov, and R.~Puzio, ``{BF description of higher-dimensional
  gravity theories},'' {\em Advances in Theoretical and Mathematical Physics}
  {\bf 3} (1999) 1289--1324, {\tt arXiv:hep-th/9901069}.

\bibitem{PeldanActionsForGravity}
P.~Peldan, ``{Actions for gravity, with generalizations: A Review},'' {\em
  Classical and Quantum Gravity} {\bf 11} (1994) 1087--1132, {\tt
  arXiv:gr-qc/9305011}.

\bibitem{MyersHigherDerivativeGravity}
R.~Myers, ``{Higher-derivative gravity, surface terms, and string theory},''
  {\em Physical Review D} {\bf 36} (1987) 392--396.

\bibitem{TeitelboimDimensionallyContinuedTopological}
C.~Teitelboim and J.~Zanelli, ``{Dimensionally continued topological
  gravitation theory in Hamiltonian form},'' {\em Classical and Quantum
  Gravity} {\bf 4} (1987) L125--L129.

\bibitem{MaedaLovelockBlackHoles}
H.~Maeda, S.~Willison, and S.~Ray, ``{Lovelock black holes with maximally
  symmetric horizons},'' {\em Classical and Quantum Gravity} {\bf 28} (2011)
  165005, {\tt arXiv:1103.4184 [gr-qc]}.

\bibitem{MyersBlackholeThermodynamicsIn}
R.~Myers and J.~Simon, ``{Black-hole thermodynamics in Lovelock gravity},''
  {\em Physical Review D} {\bf 38} (1988) 2434--2444.

\bibitem{ExirifardLovelockGravityAt}
Q.~Exirifard and M.~Sheikh-Jabbari, ``{Lovelock gravity at the crossroads of
  Palatini and metric formulations},'' {\em Physics Letters B} {\bf 661} (2008)
  158--161, {\tt arXiv:0705.1879 [hep-th]}.
  
  
\bibitem{GhoshAnImprovedEstimate}
A.~Ghosh and P.~Mitra, ``{An improved estimate of black hole entropy in the
  quantum geometry approach},'' {\em Physics Letters B} {\bf 616} (2005)
  114--117, {\tt arXiv:gr-qc/0411035}.

\bibitem{ThiemannQSD8}
T.~Thiemann, ``{Quantum spin dynamics: VIII. The master constraint},'' {\em
  Classical and Quantum Gravity} {\bf 23} (2006) 2249--2265, {\tt
  arXiv:gr-qc/0510011}.
  
  \bibitem{ThiemannQSD1}
T.~Thiemann, ``{Quantum spin dynamics (QSD)},'' {\em Classical and Quantum
  Gravity} {\bf 15} (1998) 839--873, {\tt arXiv:gr-qc/9606089}.
  
  \bibitem{DittrichTestingTheMasterII} 
  B.~Dittrich and T.~Thiemann, ``{Testing the master constraint programme for loop quantum gravity. II. Finite dimensional systems},''
 {\em Classical and Quantum Gravity} {\bf 23}, (2010) 1067--1088, {\tt arXiv:gr-qc/0411139}.
 
 \bibitem{DittrichTestingTheMasterV} 
  B.~Dittrich and T.~Thiemann, ``{Testing the master constraint programme for loop quantum gravity. V. Interacting field theories},''
  {\em Classical and Quantum Gravity} {\bf 23}, (2006) 1143--1162,
  {\tt arXiv:gr-qc/0411142}.

\bibitem{BarrettLorentzianSpinFoam}
J.~W. Barrett, R.~J. Dowdall, W.~J. Fairbairn, F.~Hellmann, and R.~Pereira,
  ``{Lorentzian spin foam amplitudes: graphical calculus and asymptotics},''
  {\em Classical and Quantum Gravity} {\bf 27} (2010) 165009, {\tt
  arXiv:0907.2440 [gr-qc]}.

\bibitem{NeimanOnShellActions}
Y.~Neiman, ``{On-shell actions with lightlike boundary data},'' {\tt
  arXiv:1212.2922 [hep-th]}.

\bibitem{BXV}
N.~Bodendorfer, ``{Black hole entropy from loop quantum gravity in higher
  dimensions},'' {\em Physics Letters B} {\bf 726} (2013) 887--891, {\tt
  arXiv:1307.5029 [gr-qc]}.

\bibitem{BekensteinTheQuantumMass}
J.~Bekenstein, ``{The quantum mass spectrum of the Kerr black hole},'' {\em
  Lett. Nuovo Cimento} {\bf 11} (1974) 467.

\bibitem{BekensteinQuantumBlackHoles}
J.~Bekenstein, ``{Quantum Black Holes as Atoms},'' in {\em Prodeedings of the
  Eight Marcel Grossmann Meeting} (T.~Piran and R.~Ruffini, eds.), (Singapore),
  pp.~92--111, World Scientific, 1999.
\newblock {\tt arXiv:gr-qc/9710076}.

\bibitem{MedvedOnTheUniversal}
A.~J.~M. Medved, ``{ON THE "UNIVERSAL" QUANTUM AREA SPECTRUM},'' {\em Modern
  Physics Letters A} {\bf 24} (2009) 2601--2609, {\tt arXiv:0906.2641 [gr-qc]}.

\bibitem{KothawalaIsGravitationalEntropy}
D.~Kothawala, T.~Padmanabhan, and S.~Sarkar, ``{Is gravitational entropy
  quantized?},'' {\em Physical Review D} {\bf 78} (2008) 104018, {\tt
  arXiv:0807.1481 [gr-qc]}.

\bibitem{BarberoFluxAreaOperator}
J.~{Barbero G.}, J.~Lewandowski, and E.~Villasenor, ``{Flux-area operator and
  black hole entropy},'' {\em Physical Review D} {\bf 80} (2009) 044016, {\tt
  arXiv:0905.3465 [gr-qc]}.

\bibitem{FreidelTwistedGeometriesA}
L.~Freidel and S.~Speziale, ``{Twisted geometries: A geometric parametrization
  of SU(2) phase space},'' {\em Physical Review D} {\bf 82} (2010) 084040, {\tt
  arXiv:1001.2748 [gr-qc]}.

\bibitem{AshtekarMechanicsOfHigher}
A.~Ashtekar, T.~Pawlowski, and C.~V.~D. Broeck, ``{Mechanics of higher
  dimensional black holes in asymptotically anti-de Sitter spacetimes},'' {\em
  Classical and Quantum Gravity} {\bf 24} (2007) 625--644, {\tt
  arXiv:gr-qc/0611049}.

\bibitem{GravanisConservedChargesIn}
E.~Gravanis, ``{Conserved charges in (Lovelock) gravity in first order
  formalism},'' {\em Physical Review D} {\bf 81} (2010) 084013, {\tt
  arXiv:1004.3582 [gr-qc]}.


\end{thebibliography}
\end{document}